# Exciton Formation in Two-Dimensional Semiconductors

K. Mourzidis[1*], V. Jindal[1*], M. Glazov[2], A. Balocchi[1], C. Robert[1], D. Lagarde[1], P. Renucci[1], L. Lombez[1], T. Taniguchi[3], K. Watanabe[4], T. Amand[1], S. Francoeur[5], X. Marie[1,6]†

[1]*Université de Toulouse, INSA-CNRS-UPS, LPCNO, 135 Av. Rangueil, 31077 Toulouse, France*
[2]*Ioffe Institute, 26 Polytechnicheskaya, 194021 Saint Petersburg, Russia*
[3]*International Center for Materials Nanoarchitectonics, National Institute for Materials Science, 1-1 Namiki, Tsukuba 305-00044, Japan*
[4]*Research Center for Functional Materials, National Institute for Materials Science, 1-1 Namiki, Tsukuba 305-00044, Japan*
[5]*RQMP and Département de génie physique, Polytechnique Montréal, Montréal, Québec H3C 3A7, Canada.*
[6] *Institut Universitaire de France, 75231 Paris, France*

*The optical properties of atomically thin semiconductors are dominated by excitons, tightly bound electron-hole pairs, which give rise to particularly rich and remarkable physics. Despite their importance, the microscopic formation mechanisms of excitons remain very poorly understood due to the complex interplay of concurrent phenomena occurring on an ultrafast timescale. Here, we investigate the exciton formation processes in 2D materials based on transition metal dichalcogenide (TMD) monolayers using a technique based on the control of excitation light polarization. It allows us to distinguish between the two competing models of exciton formation: geminate and bimolecular formation. The geminate process is the direct formation of the exciton from the initially photogenerated electron hole pair before the loss of correlation between them, whereas the bimolecular process corresponds to the random binding of free electron hole-pairs from the initially photogenerated plasma. These processes control the exciton formation time.*

*Our findings reveal that the luminescence intensity is higher by up to 40% for circularly polarized excitation compared to linearly polarized excitation for laser energy above the free carrier gap. We show that this spin-dependent exciton emission is a fingerprint of the bimolecular formation process. Importantly, we observe that exciton linear polarization (valley coherence) persists even for laser excitation energies exceeding the gap. We demonstrate that it is the result of a fraction of excitons formed by a geminate process. This shows that two formation processes coexist for excitation energies above the gap, where both mechanisms operate concurrently.*

*Similar results obtained on the two most emblematic materials of the TMDs semiconductor family, WSe$_2$ and MoS$_2$ monolayers, confirm this dual formation mechanism. These findings overturn the prevailing simplistic view of purely geminate or bimolecular exciton formation and provide crucial insights on exciton physics and means to preserve spin/valley coherence. Our methodology can be applied to a broader range of semiconductor nanostructures for investigating the intricate interplay between non-resonant excitation conditions, Coulomb interactions, electron-phonon couplings, excitonic dynamics and quantum coherence.*

*These authors contributed equally to this work.

† marie@insa-toulouse.fr

I. INTRODUCTION

Excitons, correlated electron-hole pairs bound by attractive Coulomb interaction, govern the optical response of many semiconductors and their characteristics have been extensively investigated over several decades [1–6]. Despite their critical role in defining material properties and the operation of many optoelectronic devices, the precise mechanisms of exciton formation following non-resonant light absorption remain inadequately understood for most semiconductors, including Transition Metal Dichalcogenide (TMD) monolayers characterized by very robust excitons with binding energy of hundreds of meV [7]. Very few experimental results give a direct insight in the exciton formation process, despite the great progress in the development of ultra-fast spectroscopy. Time-resolved experiments generally do not allow a clear identification of the exciton formation mechanism since there is a complicated interplay of various dynamical processes occurring at the same time on ultra-short scales (picosecond or less) : carrier-carrier interactions, emission of phonons controlling the carrier/exciton cooling, coupling to light, etc [3,8].
The current understanding of exciton formation distinguishes between two primary mechanisms: geminate and non-geminate formation, the latter also known as bimolecular formation. These two mechanisms are schematically represented in Fig. 1 and can be simply described as follows. After optical excitation, an inter-band coherent polarization is initially generated. For the geminate case, this initial coherent electronic polarization produced by the incoming photons is converted into a strongly correlated (momentum and spin) electron and hole pair. These pairs, or hot excitons, loose excess kinetic energy through multiple inelastic scattering events and then eventually populate low-energy excitonic states where recombination and photon emission can occur. In this geminate process, the electron and hole of the annihilated exciton can be traced back to the same exciting photon, thereby preserving some correlations between the incoming and outgoing photons. In contrast, the non-geminate formation results from the random bimolecular binding of independently distributed electrons and holes : the emitting exciton in this bimolecular process originates from an electron and hole created by two distinct photons, implying that all correlations were lost in the chain of events separating absorption and emission [9,10]. In both cases, the process is usually accompanied by emission of phonon, ensuring energy and wavevector conservation.

The distinction between these two processes is fundamentally important, as it determines the dynamics and coherence of excitons and the capacity to exploit their properties for both optoelectronic and quantum applications. For instance, exciton formation rates are linear with the excitation rate for the case of geminate formation, but quadratic for the case of bimolecular formation. Photon emission from the latter suffers from additional latency and jitter, as it relies on the predominantly random encounters of two independent carriers. Geminate formation provides a direct and more controllable pathway: since the electron-hole pair originates from the same photon, exciton spin can be engineered through optical orientation, thereby allowing for the deterministic control of the exciton states. Geminate formation is required for the transfer of quantum information carried by exciting photon to the exciton (spin or valley coherence), which is crucial for quantum applications and the coherent control of optical processes and quantum states [11] . Hence, geminate formation is preferred, but the conditions in which this can be realized remain to this day poorly understood or debated.

Indeed, the literature presents only few experimental results giving a direct insight into the exciton formation processes and bimolecular recombination largely dominates for most traditional semiconductors. Bimolecular formation of excitons was demonstrated early in silicon, which allowed the measurement of the bimolecular formation coefficient [12]. In GaAs quantum wells, several experimental investigations also demonstrated the key role played by the bimolecular formation process [13–21], and a square-law dependence of the exciton formation rate on the density of electron-hole pair was evidenced. On the theoretical front, calculations confirmed that the formation of excitons is bimolecular for electron-hole densities larger than $10^9$ cm$^{-2}$ [9]. The same work also revealed that geminate formation could dominate at very low carrier densities, but this was never experimentally confirmed in III-V quantum wells to our knowledge. In high-quality GaAs and CdSe nanowires, bimolecular formation of excitons from the uncorrelated electron–hole plasma was also proposed to explain the exciton dynamics [22–24]. Most interestingly, the geminate formation was most clearly evidenced in polar bulk semiconductors like CdS due to the strong electron-phonon coupling [25]. Photoluminescence excitation (PLE) and photoconductivity spectra in these materials show oscillatory behavior with a period equal to the LO phonon energy, indicating the geminate formation of excitons and their subsequent relaxation through LO phonon emission. This reveals special conditions favoring geminate formation and the capacity of some phonons to play a significant role in preserving electron-hole correlations even for non-resonant excitations.

In TMD monolayers (ML), both experimental and theoretical investigations concluded that the formation of excitons occurs over very short times, of the order of picoseconds or less [26–32]. However, the exciton formation process remains unclear and contradictory results are reported in the literature. Exciton binding energies, typically on the order of hundreds of meV [7,33], and strong exciton-phonon interactions may favor a geminate formation under a wide range of excitation conditions in these 2D semiconductors. Indeed, Robert *et al.* pointed out that the strong exciton-phonon coupling should yield a geminate exciton formation process even for high excitation energy [34]. On the basis of ultra-fast pump-probe reflectivity experiments performed in $MoS_2$ ML for above band gap excitation, Trovatello *et al* also suggested that the geminate mechanism mediated by strong exciton-phonon interaction should be the dominant process for the formation of excitons [29]. In contrast, in rather similar pump-probe experiments performed on $MoS_2$, $MoSe_2$, $WS_2$ and $WSe_2$ monolayers an exciton formation from cooled electron-hole carriers was considered; this corresponds to a non-geminate process [26]. It should be emphasized that the time-resolved experiments do not allow for a clear identification of the exciton formation process and formation time since the dynamics are strongly influenced by other relaxation processes such as carrier/exciton thermalization and cooling [9]. Also problematic is the fact that experiments usually involve rather large optical excitation power. Dense photogenerated carrier densities trigger non-linear processes (Auger recombination, two-photon absorption, interplay of excitons and electron-hole plasma, screening of Coulomb interaction…), complicating the interpretation or simply hiding the exciton formation mechanism. On the theory side, Brem et al. calculated the exciton formation and relaxation cascade in $MoSe_2$ monolayers with a fully quantum mechanical description, but the calculation was performed only for excitation energies below the gap [35]. In another calculation of the exciton formation assisted by longitudinal optical phonons in TMD monolayers, an inverse square law dependence of the exciton formation time was obtained, a clear signature of bimolecular exciton formation process [27]. The dark exciton formation was also investigated experimentally and theoretically but only the transfer from a bright photogenerated exciton population was considered [36,37]. In summary, the question whether the exciton dynamics in TMDs are dominated by (i) the direct formation (geminate) and relaxation of hot excitons or (ii) the relaxation of hot carriers with subsequent exciton formation (bimolecular) and relaxation has yet to be clarified. This important knowledge gap persists despite remarkable advances in ultrafast spectroscopy, as time-resolved experiments struggle to isolate the microscopic formation processes from the complex interplay of concurrent phenomena occurring on an ultrafast time-scale: hot carrier cooling, carrier-carrier interactions, exciton formation, cooling to the emission light cone and exciton annihilation through emission. Alternative experimental strategies must be developed to address this knowledge gap.

In this work, we present two remarkably simple continuous-wave techniques which allow for discriminating between geminate and bimolecular exciton formation processes and apply them to two exemplary TMDs, $WSe_2$ and $MoS_2$ monolayers, for excitation energies both below and above the free-particle gap.
These techniques are based on an analysis of the polarization-resolved intensities or the polarization state of the exciton emission under excitation conditions where either exciton populations or coherences are created.
For an excitation energy above the gap, we observe that the total luminescence intensity of the bright excitons is larger for circularly-polarized laser light than linearly-polarized light. We show that this spin-dependent exciton luminescence results simply from the different photo-generated populations of bright and dark excitons, as a result of the random binding of electrons and holes with different spins; for linearly-polarized excitation, a larger fraction of dark exciton is initially generated, yielding a weaker population of bright exciton and hence a smaller luminescence intensity. This spin-dependent emission is a fingerprint of bimolecular formation process. On the contrary, the total luminescence intensity for excitation energies below the gap does not depend on the laser polarization, as expected for a geminate formation process. In order to validate the technique and the methodology, we carried out systematic measurements as a function of laser energy in different monolayers by means of photoluminescence excitation spectroscopy (PLE).

In a complementary manner, we have developed another method to probe the geminate exciton formation process. It is also based on the control of the polarization of the excitation light, but this time we analyze the polarization of the exciton luminescence. Remarkably, for excitation energies above the gap, we observe that the bright exciton luminescence is linearly-polarized following linearly-polarized excitation. This linearly polarized exciton emission called optical exciton alignment (or valley coherence for TMDs) is a consequence of maintaining the correlation between the initially photogenerated electron-hole pair. This can result only from a geminate formation process because, evidently, any random binding of free electron hole-pairs from an initially photogenerated plasma cannot yield linear polarization of the emission. In contrast to previous work where valley coherence was observed only for below gap excitation [38–43], we show that the observed exciton linear polarization occurs on an excitation laser range a few hundreds of meV above the gap; this corresponds to the Sommerfeld enhancement region which is much larger in TMDs compared to other semiconductor structures, as it was theoretically predicted recently [44].

Our results therefore demonstrate that the exciton formation process is dual: the two formation mechanisms (geminate and bimolecular) occur in parallel for the typical excitation powers usually used in the optical spectroscopy experiments of 2D TMDs. This means that we must go beyond the simple description used until now of either totally geminate or totally bimolecular formation process. We develop an analytical model taking into account both processes, which is in very good agreement with the measurements.

The determination of the exciton formation process we demonstrate here in WSe$_2$ and MoS$_2$ monolayers should apply to many other semiconductor nanostructures, in which spin-valley locking effects do not occur. This includes quantum wells, perovskites, various 2D materials and moiré heterostructures [45–52].

The organization of this paper is as follows. The results on spin-dependent luminescence intensity in WSe$_2$ monolayers as a probe of the bimolecular exciton formation process are presented in Section II. The measurements on exciton linear polarization as a probe of geminate exciton formation are detailed in Sect. III. Similar investigations performed in MoS$_2$ monolayers are summarized in Sect. IV. Section V presents the exciton formation model which gives results in very good agreement with the measurements. The conclusion is given in Sect. VI.

## II. SPIN-DEPENDENT EXCITON LUMINESCENCE INTENSITY AS A PROBE OF BIMOLECULAR EXCITON FORMATION

### A. Basic principle of the technique

We first show that a clear signature of bimolecular exciton formation can be obtained by simply measuring the total photoluminescence intensity of optically active ("bright") excitons for different polarizations of the laser excitation. The technique we present here is based on the control of the spin polarization of photogenerated carriers using optical orientation.

As the principles of the technique are very general and can be applied to many semiconductor structures, we present it first in a simple situation where we consider a conduction band (CB) with electron spins either ↑ or ↓ and a valence band (VB) with missing electron (hole) spins ⇑ or ⇓. Such system corresponds to the "simple bands" model in semiconductor physics.

Let us first consider a bimolecular exciton formation process. As a result of the optical selection rules, a right ($\sigma^+$) or left ($\sigma^-$) circularly-polarized excitation yields the photogeneration of ↑⇑ and ↓⇓ electron-hole pairs respectively [53] . So, only ↑⇑ excitons are generated following $\sigma^+$ excitation. In contrast, a linearly-polarized excitation (σˣ), which can simply be viewed here as a superposition of σ⁺ and σ⁻ excitation, generates equal populations of ↑⇑, ↑⇓, ↓⇑ and ↓⇓ excitons as a result of the random binding of electrons and holes, see Fig. 2(a). Excitons with parallel spins (↑⇑ or ↓⇓) can couple to light, whereas excitons with anti-parallel spins (↑⇓ or ↓⇑) are optically inactive ("dark") as the electric dipole operator does not change the spin.

If we neglect in this first simple presentation the relaxation from bright exciton to dark exciton during the lifetime, this means that we expect to detect a total luminescence intensity twice as strong for circularly polarized excitation (σ⁺) as for linearly polarized excitation (σˣ) :

$$\frac{I_{circ}}{I_{lin}} = 2,$$

for the same incident power. In Fig. 2(a), we illustrate this process considering 4 excitation photons. As a result of the bimolecular exciton formation process, the σ⁺ (σˣ) excitation yields the radiative recombination of 4 (2) bright excitons.

Let us consider now a geminate formation process.

As schematically shown in Fig. 2(b), the circularly-polarized excitation σ⁺ yields the formation of 100% bright excitons, similarly to the bimolecular case. The key difference occurs for linearly-polarized excitation which yields the photogeneration of only bright excitons in the geminate case as well since the exciton formation occurs via a virtual state in the radiative window, see Fig. 1(c). In fact, only bright exciton can be initially photogenerated, whatever the polarization of the excitation light. As a consequence, the total exciton luminescence intensity should not depend on the polarization of the excitation light, *i.e.* we expect

$$\frac{I_{circ}}{I_{lin}} = 1$$

in the case of geminate formation process. In summary, the measurement of the exciton total luminescence intensity for different excitation laser polarization reveals precious information on the exciton formation process.

## B. Spin-dependent exciton luminescence in WSe$_2$ monolayers

To test the validity and generality of this technique, we applied it to the two most studied two-dimensional materials in the TMD family. We first present the results on WSe$_2$ monolayers and we will demonstrate in Sect. IV that the same technique also provides information on the exciton formation process in MoS$_2$ monolayers.

Figure 3(a) presents the photoluminescence spectrum following a laser excitation energy of $E_{exc}$ = 1.848 eV in a hBN encapsulated WSe$_2$ monolayer (see Fig. 1(a) and Appendix A for details on the sample). In agreement with previous reports, the luminescence is dominated by two clearly identified lines corresponding to the recombination of the bright exciton $X_0^{1s}$ (1.708 eV) and the spin-forbidden dark exciton $X_D$ (1.669 eV) [54–56]; their linewidths are 2.1 meV and 1.6 meV (Full Width at Half Maximum), attesting to the high quality of the sample. The bright (dark) exciton results from the recombination of an electron from the top (bottom) conduction band and a hole in the valence band (see Fig. 2(c) ). For the sake of simplicity, we consider here only the top valence band corresponding to A-exciton formation. We recall that the optical selection rules dictate that the $X_D$ exciton is optically forbidden for in-plane polarized light; however it can couple to z-polarized light. In the experiments presented here, the light propagates mainly along z (perpendicular to the WSe$_2$ ML plane) but we use a microscope objective with high numerical aperture (NA=0.82), yielding the detection of a fraction of this z-polarized dark/grey exciton luminescence [57,58]. As the spin forbidden dark exciton energy is lower than the energy of the bright exciton, its PL intensity is significant despite a weak oscillator strength [59]. Interestingly we observe very weak PL components, associated with the recombination of charged excitons (which should lie between the $X_0^{1s}$ and $X_D$ ) [55,60,61]. This demonstrates a vanishingly small doping of the sample and this allows us to neglect possible additional exciton formation channels assisted by resident carriers [45,62]. Figure 3(b) displays the variation of the $X_0^{1s}$ luminescence intensity as a function of the energy of the laser; this excitation of photoluminescence (PLE) spectrum, which probes the optical absorption, reveals well identified exciton excited states peaks

$X_0^{2s}$, $X_0^{3s}$ and $X_0^{4s}$, which also appear clearly in the differential reflectivity spectrum shown in Fig. 3(c) [33]. The energy gap $E_g$ (shown by the green vertical arrow) has been determined in a very similar sample by performing magneto-absorption experiments in high magnetic fields which yield a binding energy of the $X_0^{1s}$ exciton of $E_b$=167 meV [63,64].

Figure 4(a) presents the total intensity of the bright exciton $X_0^{1s}$ luminescence spectrum following a circularly-polarized or linearly-polarized excitation laser; the energy of the laser is below the gap ($E_{exc}$= 1.867 eV < $E_g$). The total intensity corresponds to the sum of right and left circularly-polarized luminescence components. We observe that the two spectra are super-imposed yielding a ratio $I_{cir}/I_{lin} = 1 \pm 0.02$. We have confirmed that this ratio is always equal to one when the energy of the excitation laser is below the gap ($X_0^{1s}$< $E_{exc}$<$E_g$). Moreover, this ratio does not change when the orientation of the linear polarization in the monolayer plane is varied (not shown). Following the discussion in section II.A (see also Fig. 2), this demonstrates that the exciton formation process is geminate. This result is expected when the excitation laser is below the gap since no free electron-hole pairs are initially photogenerated in these conditions. The oversimplified presentation of Fig. 2 considers an instant just after exciton formation. Over time, a relaxation of the bright to dark states will occur but it will affect the bright and dark populations in the same way for circular or linear excitation [65]. This explains why the ratio remains equal to 1 even for a stationary detection of the photoluminescence intensity.

Figure 4(b) shows that the same experiments performed for a laser excitation energy above the gap yield very different results. Remarkably the total luminescence intensity is larger for circular excitation compared to linear excitation (though the excitation power is strictly the same); for $E_{exc}$= 1.919 eV, we measure $I_{cir}/I_{lin} = 1.25 \pm 0.02$. This behavior was never evidenced before in TMDs to our knowledge. As shown in Fig.2(a), a luminescence ratio larger than 1 is a fingerprint of bimolecular formation process, as a consequence of the population of dark excitons following linear excitation. The simplified picture presented in Fig. 2 with a calculated ratio $I_{cir}/I_{lin}$=2 corresponds to the emission characteristics just after the laser excitation, neglecting any spin and valley relaxation. In contrast the measured ratio in Fig.4(b) corresponds to a stationary result. The model presented in Sect. V allows us to calculate the stationary luminescence ratio taking into account the relevant relaxation channels. In the reasoning carried out so far, we have assumed that 100% of the excitons were formed by a bimolecular mechanism for an excitation above the gap. As we will show in the next section, this is not the case and if a fraction of the excitons follows a geminate formation process, this will also automatically lead to a value smaller than 2. Figure 4(e) shows the variation of the ratio $I_{cir}/I_{lin}$ of the bright excitons as a function of the excitation laser energy, $E_{exc}$> $E_g$. It is typically in the range 1.5 – 1.1 up to $E_{exc}$~2.08 eV; for higher excitation energies it drops down to 1 due to the geminate formation of B-excitons which was neglected so far (note that B-excitons are responsible for the broad PLE peak at $E_{exc}$=2.13 eV, involving the second, spin-orbit split, valence band of WSe$_2$ ML).

We also measured the spin-dependent luminescence of dark excitons $X_D$. As shown in Fig. 2(a), linearly-polarized excitation yields the generation of 50% dark excitons with anti-parallel electron-hole spin in the case of strictly bimolecular formation process, whereas no dark excitons are initially photogenerated after a circularly-polarized excitation. Due to the specific band structure of TMD monolayers with the two non-equivalent valleys K$^+$ and K$^-$, the dark exciton initially created as a consequence of random binding of electron-hole pairs is an indirect exciton with the electron with spin ↑ lying in the top CB valley in K$^+$ and a hole with spin ⇓ in the K- valley. This exciton cannot couple to light in a single photon process: it is both momentum and spin-forbidden. However it was shown that the electron in the top CB can experience an efficient inter-valley spin-conserving relaxation process (through emission of a K$_3$ phonon) and ends up in the bottom CB in valley K$^-$, see Fig. 2(d) [55,66]. This process leads to the formation of the dark exciton $X_D$, which can couple to z-polarized light and is clearly visible in the luminescence spectrum in Fig. 3(a). As a consequence, the $X_D$ luminescence intensity should be a good probe of the initially formed indirect dark excitons ↑⇓.

Figure 4(c) displays the total intensity of the dark exciton $X_D$ luminescence spectrum following circularly or linearly-polarized excitation laser; the energy of the laser is below the gap ($E_{exc}$< $E_g$). The two spectra are super-imposed yielding a ratio $I_{cir}/I_{lin} \approx 1$, as expected for a geminate formation process. In Fig. 4(d), the result of the same experiment performed for a laser energy above the gap is presented. Remarkably and as expected, the $X_D$ luminescence intensity is larger for linearly-polarized excitation as more dark excitons are initially created. Note the reversal of the ratio compared to the bright exciton luminescence displayed in Fig. 4(b). For the dark exciton, the ratio is smaller than 1, as expected for a bimolecular formation process; we find $I_{cir}/I_{lin}$=0.81± 0.02. Figure 4(f) shows the variation of the ratio $I_{cir}/I_{lin}$ of the dark excitons as a function of the excitation laser energy, $E_{exc}$> $E_g$. It is typically in the range 0.75 – 0.9 up to the energy of the B excitons.

### III. EXCITON LINEAR POLARIZATION AS A PROBE OF GEMINATE FORMATION

We will now show that a complementary technique, also based on the control of the spin polarization of the carriers, can provide additional information on the formation of excitons. Let us recall that, for resonant excitation, circularly polarized light generates excitons in the basic (Bloch) states $|\pm 1 >$ ("circular excitons") and linearly X- or Y-polarized light creates excitons in the coherent superposition states $|X> = (|+1> + |-1>)/\sqrt{2}$ and $|Y> = (|+1> - |-1>)/i\sqrt{2}$ ("linear excitons"). If the spin coherence between the $|+1>$ and $|-1>$ components is preserved during the lifetime, the exciton luminescence remains linearly-polarized. This effect is known as optical alignment of excitons [25,53]. In the context of TMD monolayers with non-equivalent K valleys, the observation of linear polarization is also known as exciton valley coherence [38,39,67].
To obtain linearly polarized exciton emission, a correlation must be maintained within the initially photogenerated electron-hole pair. This can result only from a geminate formation process, as the random binding of free electron hole-pairs from an initially photogenerated plasma cannot yield linearly polarized emission because electrons and holes wavefunctions have random phases. As a consequence, linear polarization of exciton luminescence is usually observed only for strictly resonant excitation of the exciton or for excitation laser energies smaller than the free carrier gap – the conditions where the formation process of exciton is geminate [68,69]. We now demonstrate that in the TMD monolayers, a significant linear polarization of luminescence is observed, even for an excitation energy above the free carrier gap. This reveals that a fraction of the excitons is formed by a geminate process in which a correlation between the electron and the hole is maintained during the formation process. This effect which has never been reported in III-V bulk or quantum well structures results from the exceptionally strong Coulomb interactions in TMDs.

Figure 5(a) presents the variation of the linear polarization $P_{lin} = \frac{I^X - I^Y}{I^X + I^Y}$ of the neutral exciton $X_0^{1s}$ luminescence as a function of the excitation laser energy. Here, $I^X$ ($I^Y$) is the luminescence intensity co-polarized (cross-polarized) with the linearly polarized excitation laser; the PLE intensity $I^X$ co-polarized with the laser is also plotted. For excitation energies below the gap, very large linear polarization (> 60%) is observed in agreement with previous reports [38,39]. In particular we measure $P_{lin}$~65% when the laser energy is strictly resonant with the 2s exciton excited state $X_0^{2s}$, indicating that energy relaxation from $X_0^{2s}$ down to $X_0^{1s}$ preserves the linear polarization (valley coherence) of the exciton. Remarkably we also observe a significant linear polarization of the exciton luminescence for excitation energies larger than the free carrier gap $E_g$. Luminescence linear polarizations as large as 30% are measured with an oscillatory behavior that will be discussed in section V. We have carefully verified that this exciton linear polarization is not a consequence of a symmetry lowering perturbation, such as local strain or localization. Figure 5(c) displays the variation of the exciton linear polarization as a function of the laser linear polarization rotating in the ML plane: within the experimental uncertainties the exciton linear polarization follows the measured laser linear polarization. Moreover, a detailed analysis of the data allows us to rule out Raman processes as the origin of this observed polarization (see Appendix C).
The observed exciton linear polarization persists over an excitation laser range of a few hundreds of meV above the gap before vanishing. This corresponds to region where the residual effect of the Coulomb mutual attraction between an electron and a hole gives rise to a correlation of their relative position in space and an enhanced wave-function overlap, yielding a larger absorption compared to free carriers, the so-called Sommerfeld enhancement factor. Recent theoretical work predicts that this enhancement factor is particularly large in 2D materials and influences the absorption spectrum over an energy range up to ~ 400 meV above the gap for $WSe_2$ ML, a much larger value compared to other semiconductors due to stronger Coulomb effects [44]. The observation of the exciton linear polarization in the region corresponding to the Sommerfeld enhancement is perfectly consistent since linear polarization can only exist for a geminate formation process which requires to maintain a correlation between the electron and the hole.

We tested the reproducibility of the results by performing the same experiments on a very high quality WSe$_2$ charge-adjustable structure, tuned at the neutrality point. The measurements of both the spin-dependent luminescence intensity and the luminescence linear polarization confirm the results already presented for the first sample (see Appendix B).

In summary, we have demonstrated that for an excitation energy above the gap in WSe$_2$ MLs, a fraction of the excitons is formed through a bimolecular process (see section II.B) while simultaneously another fraction follows a geminate process, as evidenced by the measurement of linearly polarized luminescence. Interestingly, the measured exciton linear polarization decreases with increasing excitation laser power, as shown in Fig. 5(b) for E$_{exc}$= 1.92 eV. In the investigated power range, the PL intensity depends linearly on excitation power and Auger-like non-linear effects play a negligible role [70,71]. This power dependence will be discussed in more detail in Section V. However, the effect can be interpreted simply at this stage as being the result of a decrease in the fraction of excitons formed with a geminate process since the probability of bimolecular formation increases with increase in the excitation power unlike geminate formation which must follow a linear dependence.

**IV. EXCITON FORMATION IN MOS$_2$ MONOLAYERS**

In order to demonstrate the generality of the proposed techniques and results on the exciton formation process in TMDs, we performed the same type of experiments in MoS$_2$ monolayers. Figure 6 presents the total luminescence intensity of the bright exciton $X_0^{1s}$ following circular or linear excitation. Similarly to the WSe$_2$ monolayers, we measure I$_{cir}$/I$_{lin}$ = 1 ±0.02 for excitation energies below the free carrier gap (Fig.6(a), E$_{exc}$=2.0 eV), confirming the geminate formation of excitons. In contrast, for laser excitation energies above the gap (Fig.6 (b), E$_{exc}$=2.194 eV), the luminescence intensity depends on the polarization of the laser and we find I$_{cir}$/I$_{lin}$=1.1 ±0.02, as a consequence of a bimolecular formation process for a fraction of the photogenerated excitons. In contrast to WSe$_2$ ML, we cannot confirm the bimolecular binding of electron-hole pairs by measuring the luminescence intensity ratio of dark excitons. In MoS$_2$ MLs, the observation of the dark excitons X$_D$ which lies 14 meV below $X_0^{1s}$, requires the application of a large transverse external magnetic field [72].

The PLE spectrum detecting the $X_0^{1s}$ luminescence is displayed in Fig.6(c); the vertical arrow indicates the energy of the free carrier gap E$_g$ (the binding energy of the 1s exciton in hBN encapsulated MoS$_2$ monolayers is E$_b$=220 meV) [64]. As for WSe$_2$ MLs, we clearly observe the peaks associated with the exciton excited states $X_0^{2s}$ and $X_0^{3s}$. For an excitation energy slightly below the $X_0^{2s}$, we also note the presence of another peak corresponding to the photogeneration of B excitons ($X_B^{1s}$) involving the second valence band (the valence band spin-orbit splitting in MoS$_2$ is much smaller than that in WSe$_2$) [73,74]. We have also measured the variation of the $X_0^{1s}$ PL linear polarization as a function of the linearly-polarized excitation laser energy. Unlike most semiconductors, a clear linear polarization of the emission is observed for excitation energies above the gap as a consequence of geminate formation; this is the same observation as for WSe$_2$, seen in Fig. 5(a); we also measure a decrease of the linear polarization when the excitation power increases (Fig. 6(d)). A significant linear polarization is also observed for excitation energies below the gap, as expected. Interestingly, when the laser energy is resonant with the $X_B^{1s}$ exciton (~ 2.08 eV), the linear polarization drops almost down to zero. This shows that the hole scattering from B to A valence band induces a loss of the spin/valley coherence.

In summary the measurements in MoS$_2$ monolayers confirm the results obtained with WSe$_2$ monolayers. For excitation energies above the gap, two exciton formation processes exist in parallel: a fraction of excitons is formed by random bimolecular binding of electron-hole pairs whereas a geminate formation occurs for the other fraction.

**V. MODEL AND DISCUSSION**

To further interpret the experimental results, we have developed an analytical model considering both geminate and bimolecular formation processes. In order to simplify the notations in this section and in Fig.7, the bright ($X_0^{1s}$) and dark ($X_D$) excitons write simply B and D respectively. Figure 7 presents the schematics of the band structure and the main relaxation channels we considered in order to model the electron-hole dynamics (the relevant relaxation times are also listed in Table I). The role of phonons in the exciton formation process is described theoretically in Appendix D.

Let $g$ be the generation rate (the density of electron-hole pairs injected by the laser per unit time) and we introduce the parameter $x$ as the fraction of photogenerated electron-hole pairs which participate in a bimolecular formation process, hence $(1-x)$ of the carrier pairs bind to form excitons via the geminate process. The density of electrons in the top (bottom) CB in valley K$^+$ is denoted as $n_{+\uparrow}$ ($n_{+\downarrow}$) and the symmetric notations ($n_{-\downarrow}, n_{-\uparrow}$), are used for the populations of the CB in K$^-$ valley; $p_+$ and $p_-$ are the densities of holes in K$^+$ and K$^-$ valleys respectively; for the valence band we consider only the topmost spin sub-bands due to significant spin-orbit splitting.

For the sake of simplicity, we make the following approximations:
- We neglect any non-radiative recombination channels for free electron-hole pairs (on defects or impurities); this could be easily included in the calculation, but it does not change the conclusions on the polarization-dependent luminescence intensity.
- We also disregard electron or hole inter-valley spin relaxation time (which requires simultaneous spin and valley flips). These processes occur on typical time scales of the order of microseconds, much longer than all the other characteristic times involved here (radiative recombination, intra-valley relaxation time, etc.) [75,76].
- As the dark excitons lie a few tens of meV below the bright ones (~40 meV for WSe$_2$ ML), we consider only relaxation mechanisms from bright to dark species; the reversed processes at low temperatures are negligible.

We next analyze the two relevant cases of (i) circularly polarized and (ii) linearly polarized excitation.

*Circularly Polarized Excitation.*
Let us consider first a circularly-polarized excitation (σ+). The generation rates are written as $g_+ = g$, $g_- = 0$ and we have $p_- = 0$. Considering the bimolecular formation process, the free electron-hole pairs dynamics is governed by the following equation (the free electron-hole pairs quickly bind, forming bright excitons):

$$\frac{dn_{+\uparrow}}{dt} + \left[\gamma p_+ + \frac{1}{\tau_v} + \frac{1}{\tau_s}\right] n_{+\uparrow} = x g_+ , \qquad (1)$$

where γ (cm$^2$/s) is the bimolecular exciton formation rate. The intra-valley electron spin-flip from top to bottom conduction band is due to chiral Γ$_5$ phonons and characterized by the time constant $\tau_s$ [65,77,78]; $\tau_v$ denotes the spin-conserving inter-valley electron relaxation time (from top CB in K$^+$ valley to bottom CB in K$^-$ valley, for instance) [55]; we assume that this relaxation time also controls the lifetime of the momentum-indirect exciton $X_{-\uparrow}$, see below.

| | |
|---|---|
| $\tau_r$ | Radiative time of bright excitons |
| $\tau_d$ | Radiative time of dark excitons |
| $\tau_v$ | Inter-valley electron relaxation time (spin-conserving) |
| $\tau_s$ | Electron spin relaxation time (top to bottom CB) |
| $\tau_{sB}$ | Exciton spin relaxation time (long-range exchange interaction) |
| $\tau_x$ | Exciton non-radiative recombination time |
| $\tau_B$ | Effective decay rate of bright exciton [Eq. (6)] |

Table I. The different relaxation times considered in the model.

Under the reasonable assumption that the electron-hole dynamics is mainly governed by the binding to excitons, we obtain in the steady state from Eq. (1):

$$n_{+\uparrow}p_+ = \gamma^{-1}xg_+ \qquad (2)$$

Equation (2) is obtained in the so-called strong binding regime where

$$g \gg \frac{1}{x\gamma}\left(\frac{1}{\tau_v}+\frac{1}{\tau_s}\right)^2 \qquad (3)$$

On the other hand, the pumping should be sufficiently low to neglect exciton-exciton interaction and nonlinear processes. We expect that this criterion is fulfilled in the experimental conditions.

Thus, the bright exciton population $B_+$ in the $K^+$ valley is governed by (see Fig.7):

$$\frac{dB_+}{dt} + \left[\frac{1}{\tau_v}+\frac{1}{\tau_r}+\frac{1}{\tau_x}+\frac{1}{\tau_s}\right]B_+ + \frac{B_+ - B_-}{2\tau_{sB}} = (1-x)g_+ + x\gamma n_{+\uparrow}p_+ \qquad (4)$$

Analogous equation with $+ \leftrightarrow -$ holds for the $B_-$ exciton population in the $K^-$ valley. The exciton spin/valley relaxation time $\tau_{sB}$ of the bright excitons is due to long-range exchange interaction [79–81], $\tau_r$ is the bright exciton radiative recombination time and $\tau_x$ is the non-radiative recombination time [34,82]. The first (second) term in the right-hand side corresponds to the geminate (bimolecular) formation process. Under condition (3) and using Eq. (2), we obtain for the right-hand side $(1-x)g_+ + \gamma n_{+\uparrow}p_+ = g_+$ because, in the absence of non-radiative and spin/valley processes for free electron-hole pairs, all the absorbed photons convert to the excitons either directly (via the geminate process) or through the free electron-hole pairs (bimolecular process).

Bright excitons, in turn, feed the dark excitons characterized by populations $D_+$ and $D_-$ in the corresponding valleys due to the intra-valley electron spin relaxation with the time constant $\tau_s$. Their dynamics is given by

$$\frac{dD_\pm}{dt} + \frac{D_\pm}{\tau_x} = \frac{B_\pm}{\tau_s}, \qquad (5)$$

where, as discussed previously, we assumed that the main decay channel of the intravalley dark excitons is their non-radiative recombination, given by the time $\tau_x$. It plays a negligible role for bright excitons [83–85]. However, it governs the lifetime of dark excitons since their radiative lifetime is at least 3 orders of magnitude longer than the one for bright excitons [57,59,86].

By introducing the effective decay rate of bright excitons

$$\frac{1}{\tau_B} = \frac{1}{\tau_v}+\frac{1}{\tau_r}+\frac{1}{\tau_s}+\frac{1}{\tau_x}, \qquad (6)$$

we get the populations of bright and dark excitons in the following simple form:

$$B \equiv B_+ + B_- = \tau_B g \qquad (7)$$
$$D \equiv D_+ + D_- = B\frac{\tau_x}{\tau_s}$$

Note that the bright exciton spin-flip, controlled by the $\tau_{sB}$ constant leads to a redistribution of exciton populations between the valleys but does not affect the total population of excitonic states.

*Linearly Polarized Excitation.*

The generation rates are now written as: $g_+ = g_- = g/2$. The bimolecular process allows for formation of both intra- and intervalley excitons owing to, *e.g.*, the binding of the spin-up electron in the $K^+$ valley with the hole in the same valley or in the opposite valley $K^-$. Instead of Eq. (2), we have the following relations [under the conditions (3)]:

$$n_{+\uparrow}(p_+ + p_-) = n_{-\downarrow}(p_- + p_+) = \gamma^{-1}\frac{xg}{2}, \qquad p_+ = p_-, \qquad (8)$$

where we assumed that the intra- and intervalley exciton binding rates are equal.

Dark excitons are generated via two paths: from the bright excitons, as a consequence of the intra-valley electron spin relaxation ($\tau_s$) and from the intervalley excitons $I_{+\downarrow}$ or $I_{-\uparrow}$, via the inter-valley electron spin relaxation ($\tau_v$) (see Fig. 7). As explained in Sect. II, intervalley excitons with opposite electron and hole spin ($I_{+\downarrow}$ or $I_{-\uparrow}$) can be generated following linearly-polarized laser excitation via the bimolecular binding process. As a result, we have

$$B \equiv B_+ + B_- = \tau_B g \left(1 - \frac{x}{2}\right) \tag{9}$$
$$D \equiv D_+ + D_- = B\frac{\tau_x}{\tau_s} + \tau_x x \frac{g}{2},$$

where we assumed that the main decay process of the intervalley excitons is governed by the intervalley electron relaxation time $\tau_v$ (we have checked that the consideration of the additional intra-valley spin relaxation channel does not change significantly the results).

*Total luminescence intensity ratio for circular and linear excitation.*
The total luminescence intensity of bright ($I^B$) and dark ($I^D$) excitons under steady state conditions can be written simply:
$$I^B = \frac{B}{\tau_r} \text{ and } I^D = \frac{D}{\tau_d}, \tag{10}$$
where $\tau_r$ and $\tau_d$ are the radiative recombination time of bright and dark excitons respectively.
From equations (6) and (8), we can calculate the ratio between the total luminescence intensity following circular or linear excitation. For bright excitons and dark excitons, respectively, we obtain:
$$\frac{I^B_{cir}}{I^B_{lin}} = \left(1 - \frac{x}{2}\right)^{-1} \tag{11a}$$
$$\frac{I^D_{cir}}{I^D_{lin}} = \left(1 - \frac{x}{2} + \frac{x\tau_s}{2\tau_B}\right)^{-1} \tag{11b}$$

Naturally, if 100% of the excitons are formed by the geminate process ($x = 0$), we find as expected that both ratios are equal to 1. This is in agreement with experimental results obtained for WSe$_2$ and MoS$_2$ monolayers when the excitation laser energy is below the gap (Figs. 4 and 6).

Considering now that 100% of the excitons result from bimolecular random binding of electron-hole pairs ($x = 1$), Eq. (11a) predicts that the luminescence intensity of bright excitons will be larger for circular excitation compared to linear excitation; this is consistent with the experimental results obtained for laser energies above the gap where a fraction of excitons is generated via the bimolecular process.

For a purely bimolecular process, we find strikingly a ratio $\frac{I^B_{cir}}{I^B_{lin}} = 2$, a value identical to the one which can be very easily estimated just after the photogeneration, see Fig. 2(a).

Indeed, for $x = 1$, half of the photogenerated electron-hole pairs form bright excitons through binding within the same valley and the other half of electron-hole pairs form intervalley excitons which, under our assumptions, cannot be converted to the bright ones due to the long intervalley spin-flip time. At the same time, the ratio of the dark exciton emission under the circularly and linearly polarized emission is sensitive to the ratio of the bright exciton lifetime $\tau_B$ and the intra-valley electron spin relaxation time $\tau_s$ because this ratio determines the relative efficiency of the two channels of the dark exciton formation: from the bright states via electron spin flip or from the intervalley excitons via the spin-conserving electron valley flip.

The decay rate of bright excitons $\tau_B$ is controlled by the very short radiative lifetime [Eq. (5)]: $\tau_B \sim \tau_r \approx 2\ ps$ [34,59]. The intra-valley spin relaxation time was estimated $\tau_s \approx 10\ ps$ [65]. This yields a calculated ratio for dark excitons (x=1) : $\frac{I^D_{cir}}{I^D_{lin}} = 0.3$. This ratio smaller (larger) than 1 for dark (bright) excitons is perfectly consistent with the measurements shown in Fig. 4. However, we have assumed so far that 100% of excitons are formed by the bimolecular process. The observation of linear polarization of photoluminescence for both WSe$_2$ and MoS$_2$ monolayers for excitations above the gap clearly demonstrates that a fraction of the exciton is formed following a geminate process. This means that $0 < x < 1$.

By comparing the measured and calculated values we can make a crude estimate of the fraction $x$ of excitons formed by the bimolecular process. For WSe$_2$ ML, we measured I$_{cir}$/I$_{lin}$ =1.25 (see Fig. 4). Using equation (11a), this would correspond to x=0.4. As the excitons formed by a bimolecular process yield no linear polarization, a fraction of x=0.4 would yield a linear polarization of P$_{lin}$ ~60%. This value is larger than the one we measure in Fig. 5(a), P$_{lin}$ ~30%. This is related to the fact that we have not taken into account the relaxation of linear polarization between excitation and detection.

Although the approach is simplified, our analytical model accounts very well for the experimental results and confirms that the exciton formation for excitation energies above the gap has both a geminate and a bimolecular contribution.

Let us now discuss the PL linear polarization oscillations observed in Fig. 5(a) in WSe$_2$ ML. The photoluminescence and photoconductivity exciton spectra of various semiconductors are known to display oscillatory behaviors [25,87–91]. These oscillations originate either from the relaxation of hot carriers or hot excitons governed by phonon resonances. In contrast to prior work in other semiconductor structures, the oscillations measured in Fig.5(a) are observed mainly on the linear polarization of the 1s exciton and much less on the total luminescence intensity. We have performed a detailed analysis of these oscillation periods (see Appendix C). We measure a period of about ~19 meV for the first set of oscillations below gap and ~74 meV for the second set of oscillations appearing exclusively above gap. Remarkably, these energies correspond to the energy of one $K_1$ phonon and 2 ZO phonons ($\Gamma$ valley) respectively [55,92]. In MoS$_2$ MLs, PL linear polarization oscillations appear much less clearly in Fig. 6(c). However we observe a modulation of the polarization with a period close to E'~47 meV, which could correspond to the energy of the LO/TO phonon [93–95].

The competition between the two formation processes could explain the decrease of the measured linear polarization observed in both WSe$_2$ and MoS$_2$ MLs when the excitation density increases (Fig 5(b) and in Fig. 6(d)). Although exciton-exciton collisions can also induce a decrease in spin/valley coherence [96], it is likely that the decrease in linear polarization with excitation power reflects a decrease in the probability of geminate formation due to the increase in the bimolecular formation rate: note that the geminate formation rate varies linearly with excitation power, unlike bimolecular formation which follows a quadratic dependence.

The results of our work allow us to better understand the measurements on the formation time of excitons in different TMD monolayers. Trovatello *et al* measured an ultrafast exciton formation time (~ 10 fs) when the laser energy is close to the free carrier gap of WSe$_2$ ML, whereas they observed longer formation times for larger excitation energies [29]. In light of our results, this can be understood by a geminate formation of excitons leading to an ultrafast formation time for excitation close to the gap. For larger energies we have shown that the higher the excitation energy, the higher the probability of bimolecular formation, leading to longer effective formation times. Interestingly, field sensitive mid-infrared femtosecond experiments performed in MoS$_2$ ML (laser energy above the gap) demonstrated that ~60% of the electron-hole pairs are bound into excitons already on a sub-picosecond time scale and the remaining part evolves on the scale of several ps [28]. Though Auger-like annihilation effect might play a role in these experiments due to the large excitation density, we believe that this result is in excellent agreement with our conclusions of a fraction of excitons formed by fast geminate process and the other one following a bimolecular mechanism which occurs on slightly longer times.

We have considered in this work the two main formation mechanisms which dominate for most of the experimental conditions. We also verified that comparable results are obtained in TMD samples characterized by larger dielectric disorder and/or small residual doping. Finally, let us mention that for very large densities where a highly-nonlinear regime takes place, more complicated Auger-like processes, including trimolecular ones, could play a role in the exciton formation [97].

## VI. CONCLUSION

In conclusion, our study elucidates the dual nature of exciton formation processes in two-dimensional materials, particularly WSe$_2$ and MoS$_2$ monolayers. Through the implementation of two complementary techniques based on the control of the excitation light polarization, we have successfully differentiated between geminate and bimolecular exciton formation mechanisms. Our findings reveal that for excitation energies above the gap, a portion of excitons forms through a bimolecular mechanism, characterized by spin-dependent luminescence intensity, while another portion is formed through a geminate process, as shown by the measurement of a significant exciton linear polarization or valley coherence.

By analyzing spin-dependent luminescence intensities, we established a clear correlation between the exciton formation mechanism and the resulting exciton populations in these 2D semiconductor systems.

Furthermore, we demonstrated that the degree of valley coherence (exciton alignment) remains significant even for excitation energies several hundreds meV above the free carrier gap. This implies that the strong Coulomb interactions inherent in TMDs facilitate the maintenance of electron-hole correlations during the formation process.

Our results underscore the need to move beyond simplistic models that classify exciton formation as purely geminate or bimolecular. Instead, the coexistence of these two processes in typical experimental conditions necessitates a nuanced description of exciton dynamics, providing crucial insights for the design of quantum devices and advancing our knowledge of light-matter interactions in low-dimensional systems. Our analytical model considers both formation mechanisms and provides a framework for analyzing their relative contributions. We anticipate that our findings and methodology will pave the way for further investigations into the exciton formation dynamics in in a wide range of semiconductor nanostructures, including quantum wells, perovskites, and emerging van der Waals heterostructures.

## ACKNOWLEDGMENTS

We thank E.L. Ivchenko for fruitful discussions and L. Ren for some initial measurements . This work was supported by the Agence Nationale de la Recherche under the program ESR/EquipEx+ (Grant No. ANR-21-ESRE- 0025), the ANR projects IXTASE and the France 2030 government investment plan managed by the French National Research Agency under Grant Reference No. PEPR SPIN ANR-22-EXSP0007 (SPINMAT). S.F. thanks the invited researcher position from the EUR NanoX Grant No. ANR-17-EURE-0009 in the framework of the "Programme des Investissements d'Avenir." We also acknowledge support by the EU-funded DYNASTY project, ID: 101079179, under the Horizon Europe framework program.

## APPENDIX A: SAMPLES AND EXPERIMENTAL SET-UPS

We fabricated high-quality van der Waals heterostructure made of exfoliated $WSe_2$ or $MoS_2$ monolayers embedded in hexagonal boron nitride crystals using a dry stamping technique [98]. The $WSe_2$ or $MoS_2$ ML flakes were prepared by micromechanical cleavage of a bulk crystal (from *2D Semiconductors* or *HQ Graphene*) and transferred on the bottom layer of hexagonal boron nitride on $SiO_2$/Si substrates. Then a thin hBN layer was deposited on top of the TMD ML. The bottom hBN layer thickness, measured by atomic force microscopy, is 113 nm and 121 nm for the $WSe_2$ and $MoSe_2$ sample respectively. The top hBN layer is 15nm and 17 nm thick respectively. We also investigated the properties of the charge tunable $WSe_2$ ML device (tuned at the neutrality point), previously described in Ref. [66].

Continuous-wave micro-photoluminescence and excitation of photoluminescence (PLE) experiments were performed with a tunable super-continuum laser (linewidth ~ 2 nm); similar results were obtained with a narrow-line single-frequency tunable MixTrain Laser (SpectraPhysics). The laser spot diameter on the sample was typically ~1 μm. Unless stated all the experiments are performed with a low excitation power P= 1 μW (linear regime of excitation), corresponding to a typical photogenerated carrier density in the range $10^9$-$10^{11}$ cm$^{-2}$. Polarization measurements are performed using a combination of polarizers, quarter wave plate and half wave plate for both the excitation laser and the luminescence. The PL signal was dispersed by a spectrometer and detected by a charged-coupled device camera. For reflectivity measurements, a halogen lamp with a stabilized power supply served as the white light source.

All the experiments were performed at T= 4 K in a closed-cycle vibration-free cryostat.

## APPENDIX B: ADDITIONAL RESULTS

We applied the same experimental techniques to investigate the exciton formation on another $WSe_2$ ML sample. We measure the luminescence properties of a charge adjustable $WSe_2$ monolayer tuned at the neutrality point (the applied voltage is adjusted so that the resident carrier population is vanishingly small as attested by a very weak trion luminescence intensity) [66]. Figure 8 presents the total luminescence spectra of bright ($X_0^{1s}$) and dark ($X_D$) excitons following linear or circular excitation for excitation energy below or above the free carrier gap $E_g$. Similarly to the measurements shown in Fig. 4 on the first sample, the intensity ratio $I_{cir}/I_{lin}$ is close to 1 for excitation energies below the gap, as a result of the geminate formation process. In contrast the ratio is larger (smaller) than 1 for bright(dark) excitons when the excitation energy is above the gap; this confirms that a fraction of exciton is formed by a random bimolecular binding of electron-hole pairs in these conditions.

Figure 9 displays the variation of the luminescence linear polarization as a function of the energy of the linearly-polarized excitation laser, obtained in the same sample. These PLE data clearly evidence the linear polarization measured for excitation energies above the gap and the oscillatory behavior, already observed in the first sample (Fig. 5).

**APPENDIX C: ANALYSIS OF THE PHOTOLUMINESCENCE LINEAR POLARIZATION**

We discuss here the photoluminescence excitation of WSe$_2$ ML, where the linear polarization P$_{lin}$ of the $X_0^{1s}$ exciton is measured as function of the excitation energy E$_{exc}$. Several pronounced oscillations P$_{lin}$(E$_{exc}$) are observed for linearly polarized excitation, see Figs. 5 (b) and 9. At the same time, the emission intensity does not demonstrate any pronounced oscillations as a function of E$_{exc}$.

We performed a detailed analysis of the data to rule out an interpretation of the oscillations originating from possible Raman lines in the PLE spectra. Figure 10 displays the peaks of linear polarization extracted from the measurements presented in Fig. 5 (the abscissa corresponds to the peak number starting from the low energy region, below the gap). We observe two very different slopes for below and above gap excitation. For E$_{exc}$ <E$_g$ (small disks), PL linear polarization peaks are observed every ~19 meV, whereas for E$_{exc}$ >E$_g$ (large disks), the oscillation period is ~76 meV. Remarkably, this corresponds, in the former case, to the energy of K$_1$ phonon ($\hbar\Omega_{k1}$ ~18 meV) and, in the latter case, to twice the energy of ZO phonons ($\hbar\Omega_{k1}$ ~38 meV) [55,92]. It is interesting to note that Z-polarized single-phonon interaction with charge carriers and excitons is symmetry forbidden, while a two- ZO phonon scattering is, in general, possible. Figure 11(a) presents a color plot of the PL intensity (logarithmic scale on 5 orders of magnitude) as a function of excitation and detection energy. The corresponding PL linear polarization is displayed in Fig. 11(b). Focusing on resonance #13 at 1.988 eV, no variation of intensity can be observed in the PL intensity (see Fig. 11(a) ), while a slight increase in the PL linear polarization is detected (see the crossing between the diagonal dashed white line and the horizontal dotted line corresponding to the $X_0^{1s}$ energy in Fig. 11(b). To demonstrate that this weak Raman resonance cannot explain the linear polarization peak, we performed a fit of the emission line, taking into account a Lorentzian contribution due to the luminescence of the $X_0^{1s}$ exciton and a Raman contribution (well fitted by a Gaussian line whose width is that of the supercontinuum laser) for the different laser energies; the latter follows the variation of the laser energy unlike the PL emission. Three examples of fit results on co-polarized emission spectra are displayed for the resonance #13 in Fig.12(a). We see that excellent fits are obtained with a very small contribution of the Raman effect. Figure12(b) summarizes the results of the fit for this resonance #13 showing that the Raman contributions is more than 15 times weaker than the PL ones. We performed the same analysis on all the other peaks of Fig.10 (not shown), which demonstrate that the PL linear polarization oscillations cannot be explained by Raman like resonances. Therefore, we interpret the oscillations of the PL linear polarization as resulting from phonon-cascade-like processes of hot geminate excitons.

For particular excitation energies, E$_{exc}$ = E$_{1s}$ + N$\hbar\Omega$, *i.e.*, where the difference between the detection and excitation energy equals to the N = 1, 2,... phonon energies $\hbar\Omega$, the formation/relaxation rate becomes strongly enhanced. As a result, the polarization can be enhanced as well since the polarization degree is controlled by the ratio of the energy relaxation time $\tau_\epsilon$ and the depolarization time $\tau_d$ as [53]:

$$P \sim exp\left(-\int_0^\varepsilon \frac{d\epsilon}{\epsilon} \frac{\tau_\varepsilon}{\tau_d}\right) \tag{12}$$

Similar effect of spin polarization oscillation was observed in GaSb as a consequence of short energy relaxation time $\tau_\epsilon$ compared to the non-radiative recombination time [99]. In TMD MLs, we believe that this condition is also fulfilled due to strong exciton-phonon coupling [55]. Thus, there is basically no loss of excitons in the course of their energy relaxation. This explains the very weak oscillations observed for the luminescence intensity.

**APPENDIX D : ROLE OF THE PHONONS IN THE EXCITON FORMATION**

It is instructive to analyze the processes of the exciton formation from the theoretical standpoint considering a crystal excited by an optical pulse. Let $E(t)e^{-i\omega_c t}$ be the positive frequency component of the complex amplitude of incident field with $\omega_c$ being the carrier frequency and $E(t)$ being the pulse envelope. We neglect the wavevector of radiation for simplicity.

*D1. Excitation below the bandgap. Geminate process*

We first consider the situation where $E_{exc} < E_g$ and focus on the case where $1s$ exciton is virtually formed as a result of the driving of the TMD ML by the electromagnetic field. Due to the mismatch of the exciton and photon dispersions for a non-resonant excitation the coupling with the crystalline lattice vibrations or defects is needed to fulfil both the energy and momentum conservation laws. We consider the case of exciton-phonon interaction for specificity. Hence, as a result of the exciton-phonon interaction the phonon is emitted and the exciton scatters to the real state on its dispersion curve.

In this scenario, the wavefunction of the TMD can be represented as [100]

$$\Psi(t) = |0\rangle + B_0(t)|1s,0\rangle + \sum_{\bm{k}} C_{\bm{k}}(t)|1s,\bm{k}\rangle \otimes |ph,-\bm{k}\rangle. \qquad (13)$$

Here $|0\rangle$ is the wavefunction of the crystal in the ground state, $|1s,\bm{k}\rangle$ is the wavefunction of the crystal where the $1s$ exciton in the state with the wavevector $\bm{k}$ is excited (for $\bm{k} = 0$ we denote this wavefunction as $|1s,0\rangle$), the wavefunction of the phonon with the momentum $\bm{q}$ is written as $|ph,\bm{q}\rangle$, and we consider only one phonon mode for simplicity. The coefficients $B_0(t)$ and $C_{\bm{k}}(t)$ satisfy the set of coupled equations (see Refs. [35,100] for a more advanced modeling)

$$i\dot{B}_0(t) = (\omega_0 - i\gamma_0)B_0(t) - dE(t)e^{-i\omega_c t} + \sum_{\bm{k}} M_{\bm{k}}^* C_{\bm{k}}(t), \qquad (14a)$$
$$i\dot{C}_{\bm{k}}(t) = (\omega_{\bm{k}} + \Omega_{\bm{k}} - i\gamma_{\bm{k}})C_{\bm{k}}(t) + M_{\bm{k}} B_0(t). \qquad (14b)$$

where $\omega_{\bm{k}}$ is the exciton dispersion (reckoned from the ground state, i.e., $\omega_0 \equiv \omega_{\bm{k}=0}$ is the exciton transition frequency), $\Omega_{\bm{k}}$ is the phonon dispersion, $\gamma_{\bm{k}}$ is the damping of the corresponding state, and $d \equiv d_{1s}$ is the dipole matrix element of the $1s$ exciton. The set of Eqs. (14) is readily solved by means of the Fourier transform with the result

$$B_0(t) = -de^{-i\omega_c t} \int_{-\infty}^{\infty} \frac{d\omega}{2\pi} E_\omega \frac{e^{-i\omega t}}{\omega - \delta_0 - \sum_{\bm{k}} \frac{|M_{\bm{k}}|^2}{\omega - \delta_{\bm{k}} - \Omega_{\bm{k}} + i\gamma_{\bm{k}}}}, \qquad (15a)$$

$$C_{\bm{k}}(t) = -dM_{\bm{k}} e^{-i\omega_c t} \int_{-\infty}^{\infty} \frac{d\omega}{2\pi} E_\omega \frac{e^{-i\omega t}}{\left(\omega - \delta_0 - \sum_{\bm{k}} \frac{|M_{\bm{k}}|^2}{\omega - \delta_{\bm{k}} - \Omega_{\bm{k}} + i\gamma_{\bm{k}}}\right)(\omega - \delta_{\bm{k}} + i\gamma_{\bm{k}})}. \qquad (15b)$$

Here $E_\omega$ is the Fourier transform of the pulse envelope function and $\delta_{\bm{k}} = \omega_{\bm{k}} - \omega_c$ is the detuning between the exciton and the excitation pulse carrier frequency. Note that any correlations between excitons and phonons are quickly lost because of the phonon decay and scattering, thus the coefficient $|C_{\bm{k}}|^2$ basically represents the "incoherent" exciton population. Under reasonable condition of sufficiently narrow incident pulse with the frequency spread

$$\Delta\omega \ll \omega_c - \omega_0, \quad \omega_c - \omega_0 > 0, \quad \hbar\omega_c < E_g, \qquad (16)$$

which means that the "resonant" excitation of the $1s$ state is suppressed, the denominator writes

$$\omega - \delta_0 - \sum_k \frac{|M_k|^2}{\omega - \delta_k - \Omega_k + i\gamma_k} \approx \omega_c - \omega_0.$$

As a result, the $B_0(t) = -E(t)e^{-i\omega_c t}d/(\omega_c - \omega_0)$ follows the incident radiation waveform, while $C_k \propto e^{-i\omega_c t} \int d\omega\, E_\omega e^{-i\omega t}[\omega - \delta_k + i\gamma_k]^{-1}$ rises during the incident pulse duration $\Delta\omega^{-1}$. It corresponds to almost instantaneous exciton formation via the geminate process.

### *D2. Excitation above the band gap. Interplay of the bimolecular and geminate processes*

In the case of the above gap excitation where $\hbar\omega_c > E_g$, in addition to the processes described above (which may now involve all the bound exciton states $ns$, $n = 1,2,...$ as intermediate virtual states) and leading to the geminate exciton formation, a resonant excitation of continuum states of electron-hole pairs is possible. The fate of the photocreated electron-hole pair is controlled by the exciton-phonon interaction: The geminate electron and hole photocreated after a given photon absorption can be scattered independently by the phonons and, hence, loose their coherence sufficiently fast to bind (emitting phonons) with other carriers originating formed from other photons. It is the most likely scenario since there is a continuum of electron-hole relative motion states available for the scattering. In this scenario the exciton formation occurs with a delay related to the scattering between the real intermediate states, energy relaxation, and binding.

The initial process of the coherent photoexcitation of the TMD and formation of electron-hole pairs in the continuum states can be described along the same lines as above considering the generalized form of the wavefunction

$$\Psi(t) = |0\rangle + \sum_\nu B_{\nu,0}(t)|\nu, 0\rangle + \sum_{k,\nu'} C_{\nu',k}(t)|\nu', k\rangle \otimes |ph, -k\rangle, \tag{17}$$

where the subscripts $\nu, \nu'$ enumerate the relative motion states of the excitons (generally including both the discrete states and the continuum, correspondingly, the sums over $\nu, \nu'$ should be understood as the sums over the discrete states and integrals over the continua), $B_{\nu,0}(t), C_{\nu',k}(t)$ are the coefficients, and $k$ is the center of mass wavevector of the exciton. As before, the phonon momentum is neglected. Note that Eq. (17) describes the state of the crystal with one electron-hole pair excited. Note that in this limit of a single electron-hole pair excited in the undoped crystal, the exciton formation is geminate regardless of the energy of the absorbed photon since there are no other charge carriers to bind with. Bimolecular formation of excitons, naturally, requires formation of more than one electron-hole pair which requires absorption of more than one photon. These effects, as we see below, can be naturally accounted for using the kinetic equation approach.

The coefficient $B_{\nu,0}(t)$ can be readily found in the form

$$B_{\nu,0}(t) = -d_\nu e^{-i\omega_c t} \int_{-\infty}^{\infty} \frac{d\omega}{2\pi} E_\omega \frac{e^{-i\omega t}}{\omega - \delta_{\nu,0} + i\gamma_{\nu,0}}, \tag{18}$$

Here $d_\nu$ is the dipole matrix element which includes the Sommerfeld enhancement, $\delta_{\nu,0} = \omega_{\nu,0} - \omega_c$ is the detuning between the electron-hole pair excitation frequency and the pump pulse carrier frequency and $\gamma_{\nu,0} = \pi \sum_{\nu',k'} |M_{\nu'k';\nu,0}|^2 \delta(\omega - \delta_{\nu',k'} - \Omega_k)$ is the damping of the electron-hole pair related to its scattering to other states. Unlike the results for the below-gap excitation, the resonant absorption takes place at $\omega + \omega_c = \omega_{\nu,0}$. After the pump pulse is over, the occupancies of the continuum states are evaluated as

$$N_{\nu,0}^{(0)} = \frac{\pi}{\gamma_{\nu,0}} \int_{-\infty}^{\infty} \frac{d\omega}{2\pi} |d_\nu E_\omega|^2 \delta(\omega - \delta_{\nu,0}), \tag{19}$$

which corresponds to the Fermi golden rule result. Moreover, after a single electron-hole pair scattering by a phonon or defect the Coulomb correlations are lost, and the populations of $\boldsymbol{k} = 0$ states are distributed over the range of the center of mass wavevectors. Hence, to determine further dynamics of the system it is sufficient to apply the kinetic equation formalism. In the most simplistic approach we represent the population of continuum states as populations of uncorrelated electron-hole pairs with the electron and hole densities being

$$n(t=0) = p(t=0) \equiv \mathcal{S} \sum_{v,k} N_{v,0}^{(0)}. \qquad (20)$$

Here $\mathcal{S}$ is the normalization area. The temporal dynamics of $n(t)$ and $p(t)$ is described by the equations presented in the main text. The excitons in this regime appear with a delay related to the formation of a bound state from unbound pair.

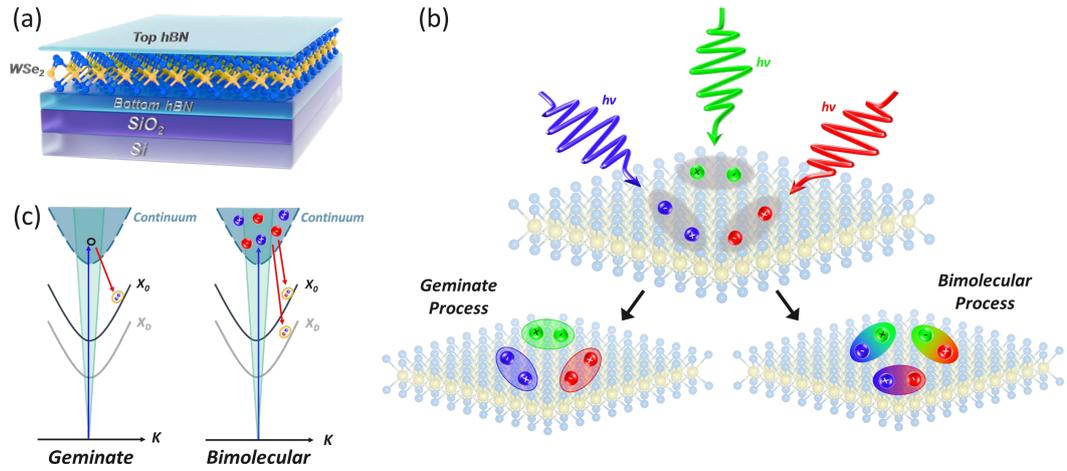

**Figure 1**. (a) Schematics of the investigated WSe$_2$ monolayer embedded in hexagonal boron nitride. Simplified illustration of the two exciton formation processes in (b) real space and (c) reciprocal space. Colors in (b) are used to identify electrons and holes created by a given photon. In (c), the parabolas correspond to the free carrier continuum (dashed), the bright (X$_0$) and dark (X$_D$) excitons (black and grey parabolas, respectively); the blue arrows represent the laser excitation (the shaded region is the radiative cone) and the red arrows represent the phonons. In the geminate process, the photogenerated electron and hole maintain mutual correlation until their binding into an exciton. In the bimolecular case the excitons are created through random binding of electrons and holes placed in the space independently of each other.

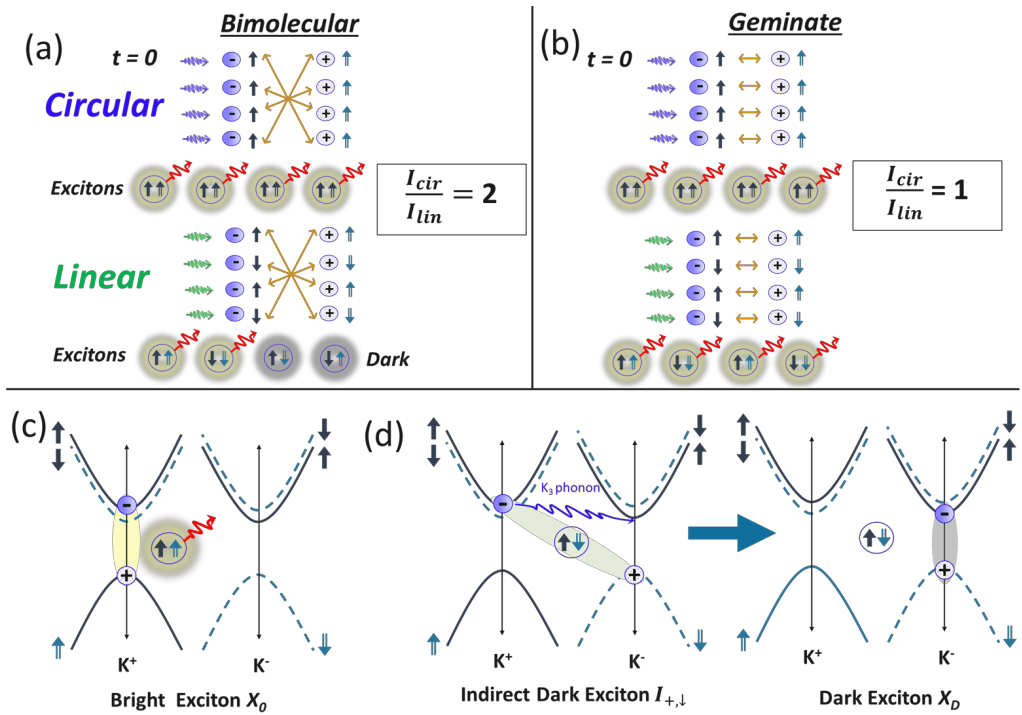

**Figure 2**. Schematic illustration of the (a) bimolecular and (b) geminate exciton formation processes for circularly or linearly polarized excitation light. For the bimolecular process, the right ($\sigma^+$) circularly-polarized excitation generates at $t = 0$ ↑⇑ electron-hole pairs which bind into ↑⇑ excitons; in contrast the linearly-polarized excitation photogenerates equal population of ↑⇑, ↑⇓, ↓⇑ and ↓⇓ excitons as a result of the random binding of electrons and holes, *i.e.* half of these excitons (↑⇓ or ↓⇑) are -optically inactive (dark). Just after the photogeneration, the total luminescence intensity is expected to be twice as strong for circularly polarized excitation as for linearly polarized excitation : $\frac{I_{circ}}{I_{lin}} = 2$. On the contrary, the same reasoning applied to the geminate formation leads to a luminescence intensity which does not depend on the polarization of the excitation: $\frac{I_{circ}}{I_{lin}} = 1$. (c,d) Band structure of a WSe$_2$ monolayer (only the top valence band A is considered) showing (c) the bright exciton X$_0$ recombination in the K$^+$ valley and (d) the intervalley dark exciton $I_{-,\uparrow}$, which can relax to the spin-forbidden X$_D$ dark exciton through an inter-valley K$_3$ phonon emission.

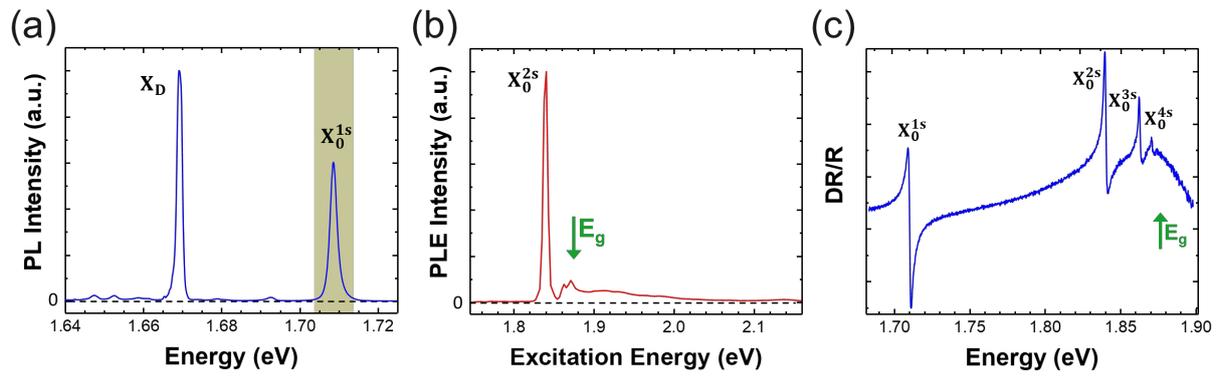

**Figure 3**. - WSe$_2$ monolayer - (a) Photoluminescence spectrum (blue curve) showing the bright exciton $X_0^{1s}$ and the spin forbidden dark exciton $X_D$; the excitation energy is $E_{exc}$= 1.848 eV and the laser is linearly polarized. (b) PLE spectrum : variation of the luminescence intensity of the bright exciton $X_0^{1s}$ (shaded area in Fig. 3(a) ) as a function of the exciton laser energy for a fixed excitation power of 1 μW. (c) Differential reflectivity spectrum (DR/R) showing the different exciton excited states; $E_g$ is the free carrier gap (see text for details).

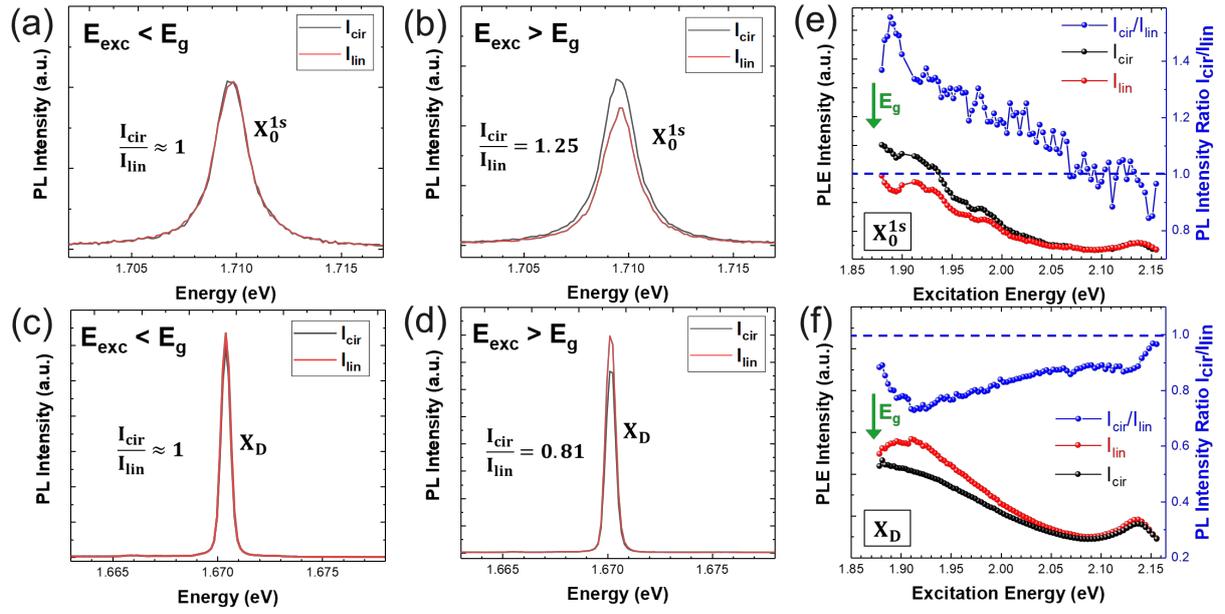

**Figure 4**. - WSe$_2$ monolayer - Polarization dependent luminescence intensity. Total luminescence spectrum of bright exciton $X_0^{1s}$ following linear or circular excitation for (a) excitation energy below the free carrier gap $E_g$ ($E_{exc}$=1.867 eV), and (b) above $E_g$ ($E_{exc}$=1.919 eV). The ratio $I_{cir}/I_{lin}$ corresponds to the spectrally integrated total luminescence intensity ratio. (c) and (d) correspond to the same type of measurement for the dark exciton $X_D$. Note the opposite behavior in terms of relative intensity for bright and dark excitons [compare (b) and (d)]. PLE spectra of (e) bright, (f) dark excitons for excitation energies above the gap for both circular and linear excitation; the corresponding intensity ratio is also plotted as a function of excitation energy.

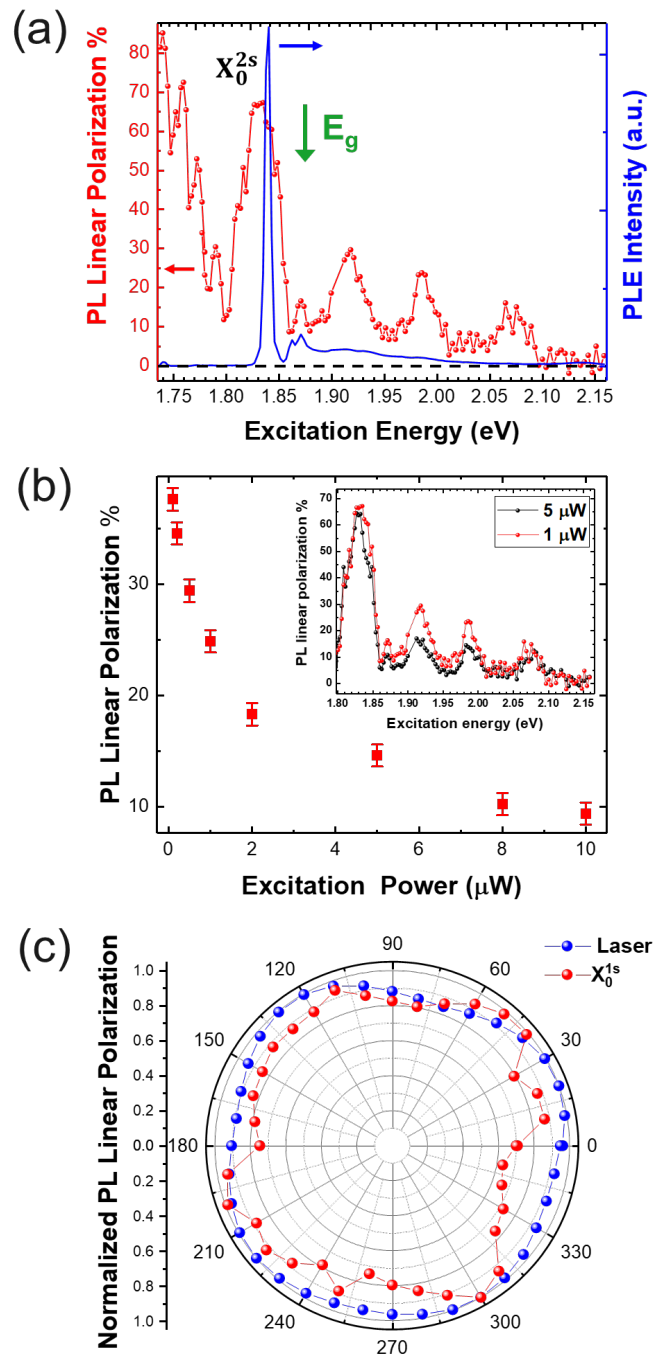

**Figure 5**. - WSe$_2$ monolayer - Valley coherence. (a) Luminescence linear polarization of the bright exciton $X_0^{1s}$ as a function of the energy of the linearly-polarized excitation laser. The blue line displays the variation of the co-polarized luminescence intensity (PLE). (b) Variation of the PL linear polarization of the bright excitons $X_0^{1s}$ as a function of the excitation power (E$_{exc}$=1.92 eV). The inset in (b) shows the PL linear polarization as a function of the excitation energy for two different excitation powers. (c) Normalized angle dependent polar plot of the laser and the bright excitons $X_0^{1s}$ luminescence linear polarization.

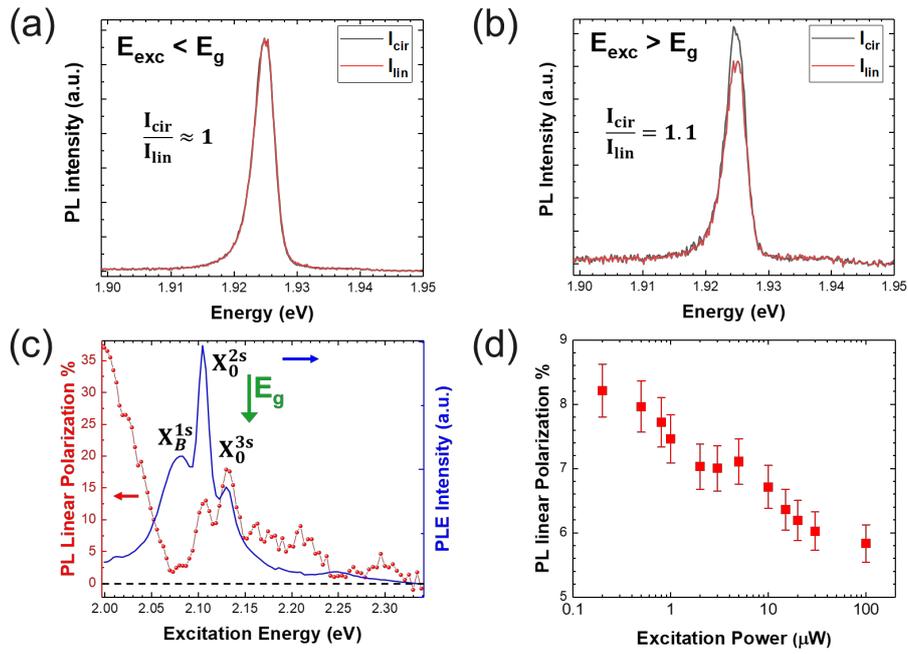

**Figure 6**. - MoS$_2$ monolayer - (a) Total luminescence spectrum of bright exciton $X_0^{1s}$ following linear or circular excitation for excitation energy (a) below the free carrier gap $E_g$ ($E_{exc}$=2.00 eV) and (b) above the free carrier gap $E_g$ ($E_{exc}$=2.195 eV). (c) Luminescence linear polarization of the bright exciton $X_0^{1s}$ as a function of the energy of the linearly-polarized excitation laser. The blue line displays the variation of the co-polarized luminescence intensity (PLE); $X_B^{1s}$ is the ground exciton state involving the second B valence band. (d) Variation of the PL linear polarization of the bright exciton $X_0^{1s}$ as a function of excitation power ($E_{exc}$=2.21 eV).

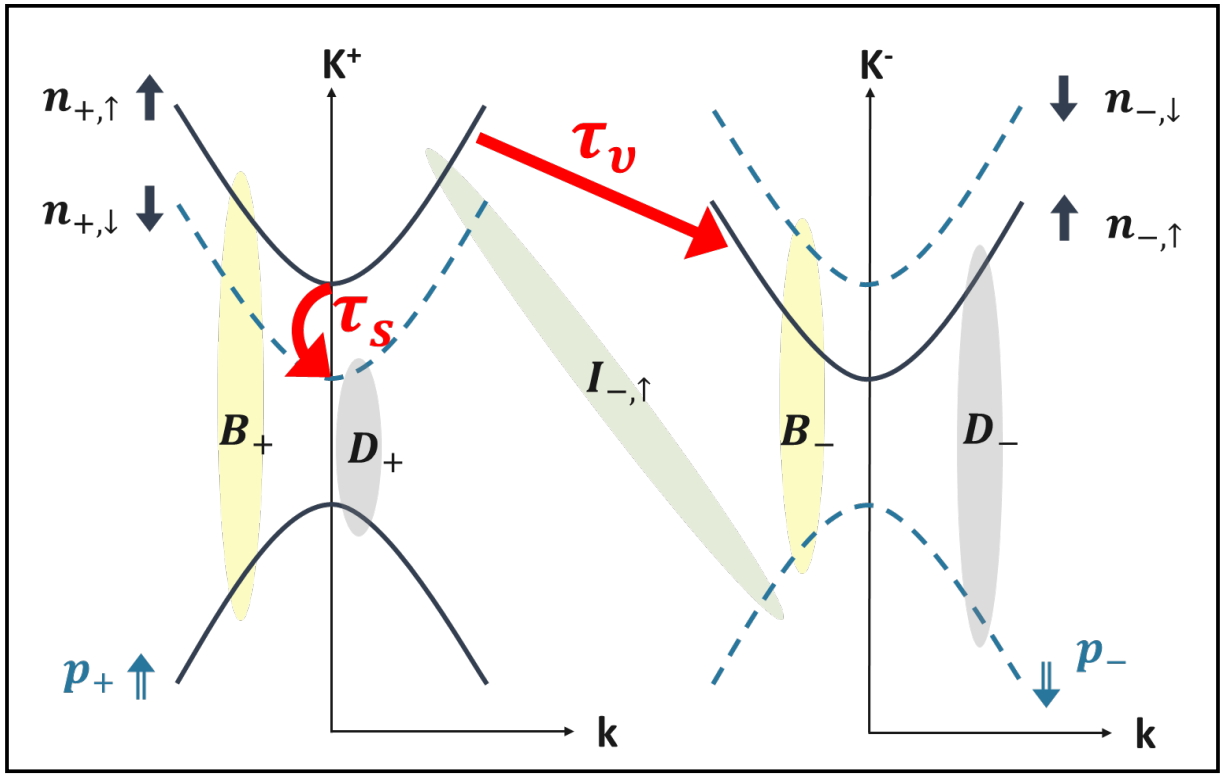

**Figure 7.** Schematic illustration of the WSe$_2$ monolayer band structure showing the bright excitons (B$_+$ and B$_-$), the dark excitons (D$_+$ and D$_-$) and the inter-valley exciton $I_{-,\uparrow}$ considered in the model. The relaxation processes described by the time constants $\tau_s$ (the intra-valley spin relaxation) and $\tau_v$ (the spin-conserving inter-valley relaxation time) are also shown.

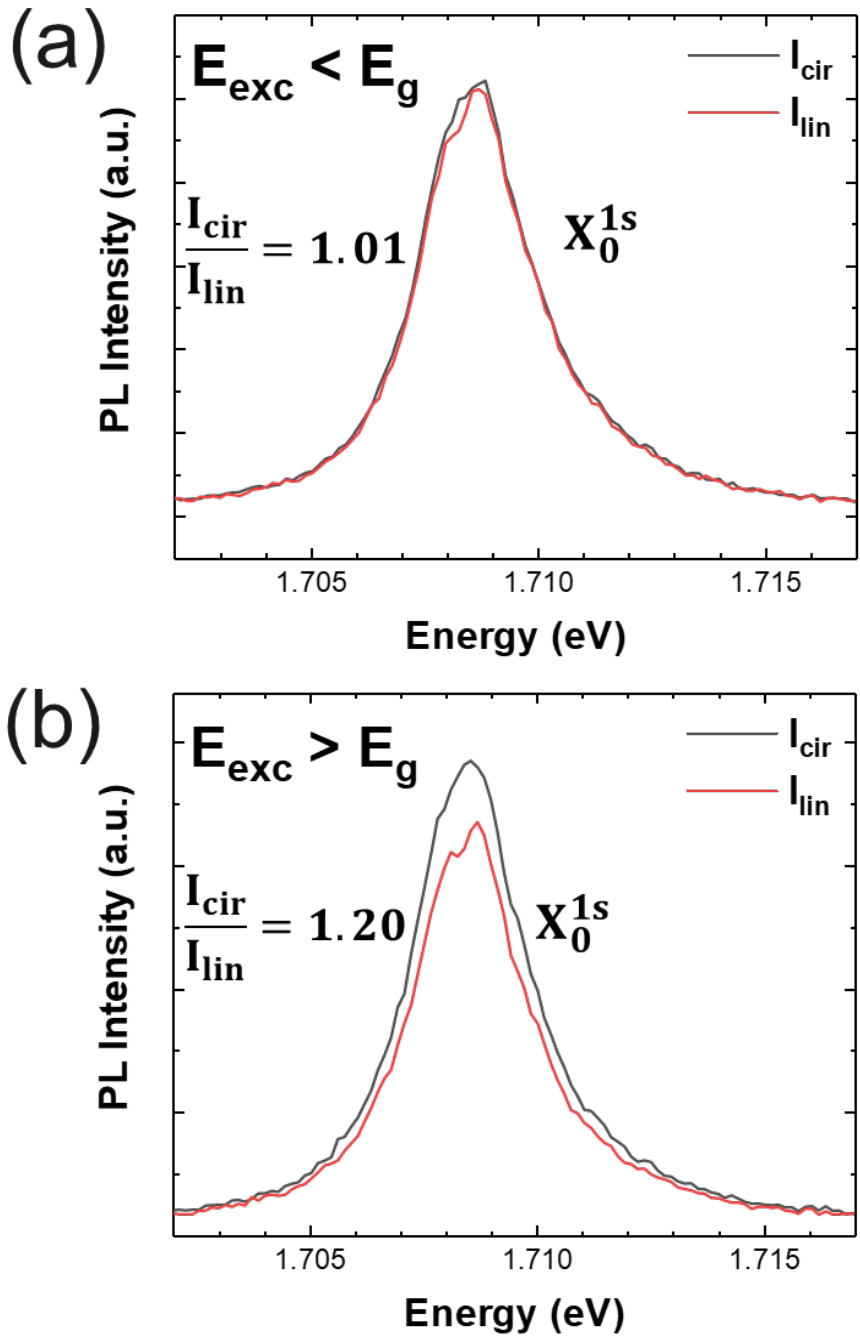

**Figure 8**. - WSe$_2$ monolayer - Polarization-dependent luminescence in a charge adjustable WSe$_2$ monolayer. Total luminescence spectra of the bright exciton $X_0^{1s}$ following linear or circular excitation for excitation energy (a) below the free carrier gap $E_g$ ($E_{exc}$=1.879 eV), and (b) above the free carrier gap $E_g$ ($E_{exc}$=1.893 eV). The ratio $I_{cir}/I_{lin}$ corresponds to the spectrally integrated total luminescence intensity ratio.

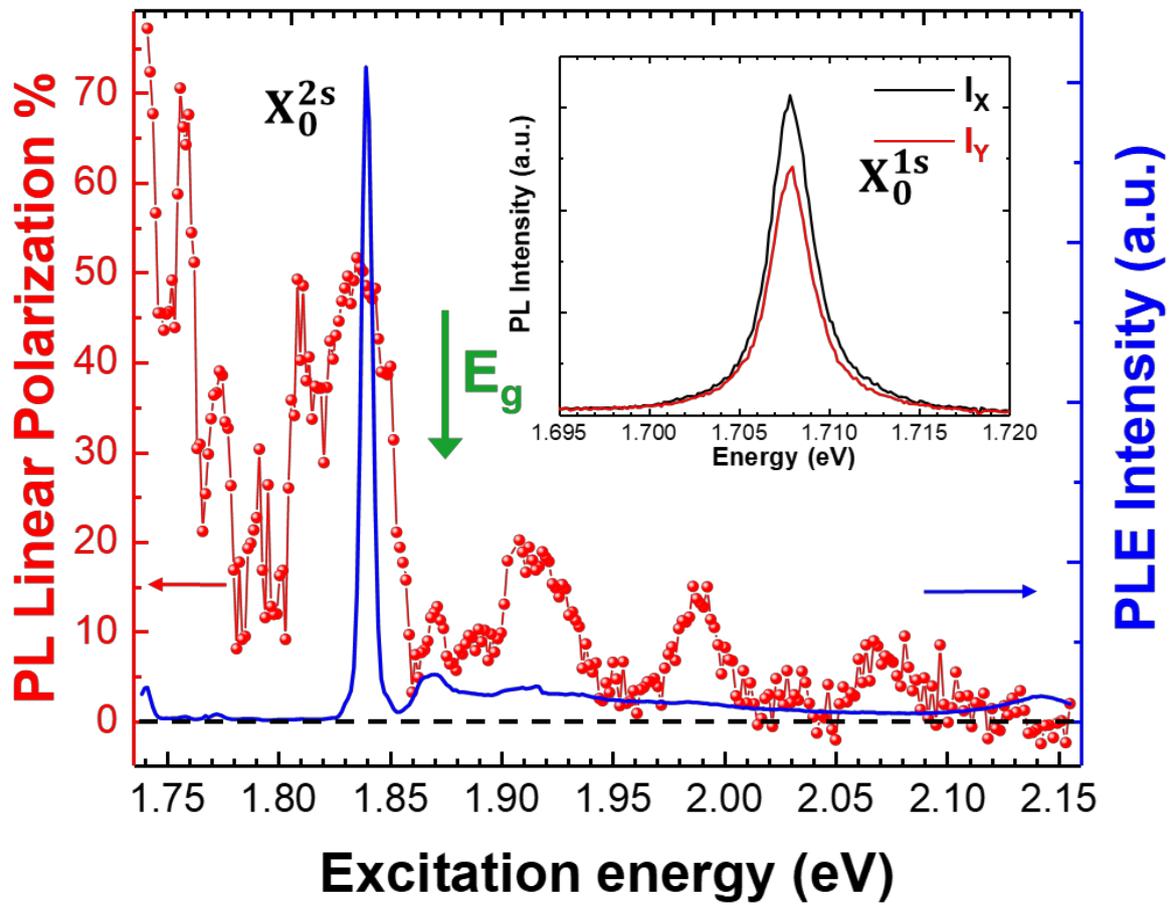

**Figure 9**. - WSe$_2$ monolayer - Valley coherence in a charge adjustable WSe$_2$ monolayer. Luminescence linear polarization of the bright exciton $X_0^{1s}$ as a function of the energy of the linearly-polarized excitation laser. The blue line displays the variation of the co-polarized luminescence intensity (PLE). The inset shows the linear co- ($I_X$) and cross- ($I_Y$) polarized components of the bright exciton photoluminescence intensities following a linearly polarized excitation (E$_{exc}$= 1.93 eV).

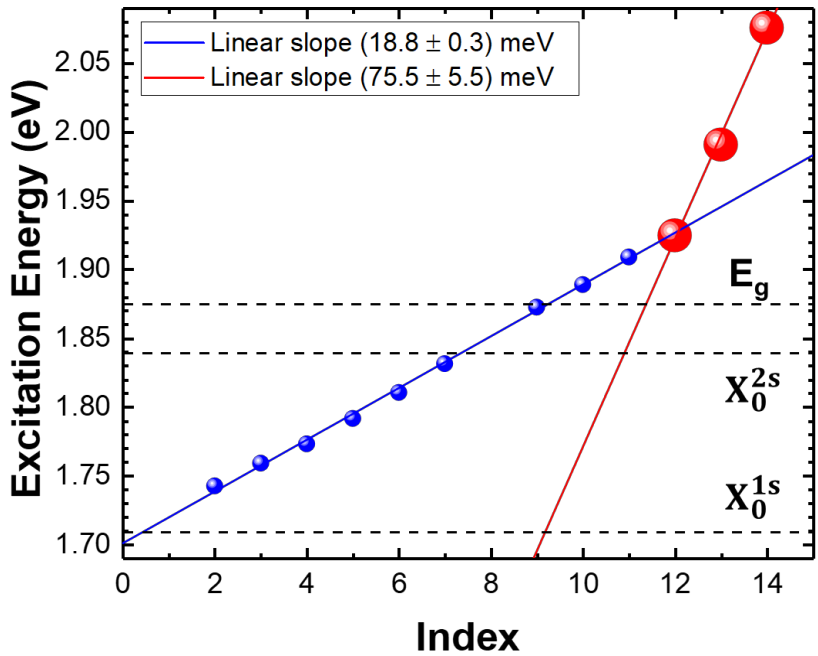

**Figure 10**. - WSe$_2$ monolayer - Analysis of linear polarization oscillations. Energy of the linear polarization peaks extracted from the measurements presented in Fig. 5 in the WSe$_2$ monolayer (the abscissa corresponds to the peak number starting from the low energy region, below the free carrier gap). For hν<E$_g$ (small disks), PL linear polarization peaks are observed every ≈18 meV, whereas for hν>E$_g$ (large disks), the oscillation period is ≈76 meV (the full lines are linear fits). The dashed horizontal lines correspond to the energy of the free carrier gap E$_g$ and the bright exciton ground ($X_0^{1s}$) and first excited ($X_0^{2s}$) state.

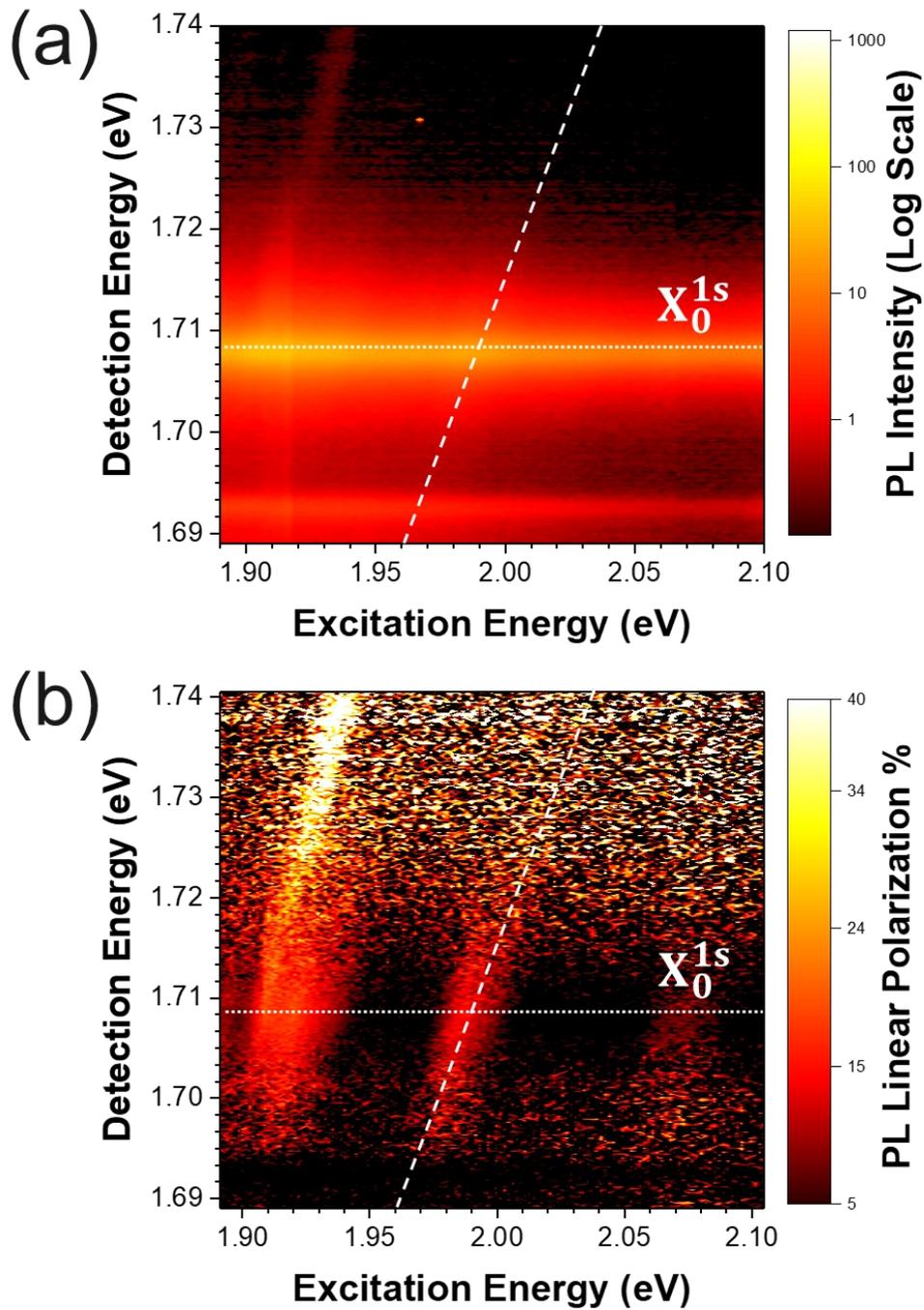

**Figure 11**. - WSe$_2$ monolayer - Two-dimensional photoluminescence analysis. (a) Color plot of the PL intensity (logarithmic scale) as a function of excitation and detection energies for resonance #13. The dotted white horizontal line indicates the emission energy of the bright exciton $X_0^{1s}$ and the diagonal dashed white line corresponds to the energy of resonance #13 (see Appendix C) (b) Color plot of the corresponding PL linear polarization.

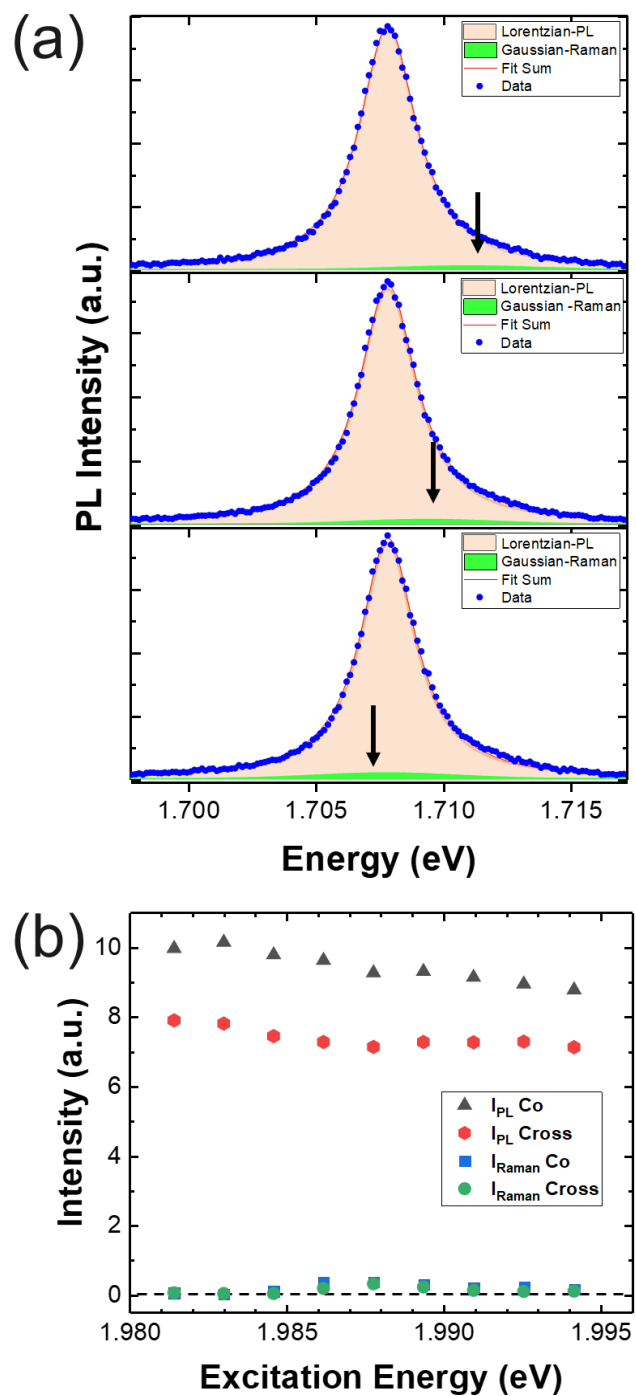

**Figure 12**. - WSe$_2$ monolayer - Distinguishing photoluminescence from Raman signals. (a) PL spectrum of the bright exciton for three different excitation energies $E_{exc}$= 1.9909, 1.9893 and 1.9877 eV close to the resonance #13. The blue dots are the experimental data while the orange and green area correspond to the PL and Raman contributions of the fit; the vertical black arrow indicates the peak of the small Raman contribution, which follows the variation of the laser energy unlike the PL emission. (b) Corresponding integrated intensities of the different contributions for co-and cross-polarized emission.